\documentclass[%
preprint,
 amsmath,amssymb,
 aps,
]{revtex4-2}
\usepackage{tabularx}
\usepackage{graphicx}
\usepackage{dcolumn}
\usepackage{bm}
\usepackage{lineno}
\usepackage{xcolor}

\begin{document}

\title{\textbf{Anharmonicity Driven by Vacancy Ordering Unlocks High-performance Thermoelectric Conversion in Defective Chalcopyrites II-III$_2$-VI$_4$}}%
\author{Hui Zhang$^{1,\dag}$, Jincheng Yue$^{2,3,1,\dag,*}$, Jiongzhi Zheng$^{4,5}$, Ning Wang$^{1}$, Wenling Ren$^{6}$, Shuyao Lin$^{7,8}$, Chen Shen$^{9,10}$, Hao Gao$^{3}$, Yanhui Liu$^{1,*}$, Yue-Wen Fang$^{2,11,*}$, Tian Cui$^{1,*}$} 

\affiliation{$^{1}$ Institute of High-Pressure Physics, School of Physical Science and Technology, Ningbo University, Ningbo 315211, China}

\affiliation{$^{2}$ Centro de Física de Materiales (CFM-MPC), CSIC-UPV/EHU, Manuel de Lardizabal Pasealekua 5, 20018 Donostia/San Sebastián, Spain}

\affiliation{$^{3}$ Fisika Aplikatua Saila, Gipuzkoako Ingeniaritza Eskola, University of the Basque Country (UPV/EHU), Europa Plaza 1, 20018 Donostia/San Sebastián, Spain}

\affiliation{$^{4}$ Energy Technologies Area, Lawrence Berkeley National Laboratory, Berkeley, CA 94720, USA}

\affiliation{$^{5}$ Thayer School of Engineering, Dartmouth College, Hanover, New Hampshire, 03755, USA}

\affiliation{$^{6}$ Institute of Materials Science, Technical University of Darmstadt, Alarich-Weiss-Strasse 2, Darmstadt, 64287, Germany}

\affiliation{$^{7}$ Technische Universit\"{a}t Wien, Institute of Materials Science and Technology, Vienna, A-1060, Austria}

\affiliation{$^{8}$ Link\"{o}ping University, Department of Physics, Chemistry, and Biology (IFM), Link\"{o}ping, SE-58183, Sweden}

\affiliation{$^{9}$ Suzhou Laboratory, Suzhou, China}

\affiliation{$^{10}$ Gusu Laboratory of Materials, Suzhou, China}

\affiliation{$^{11}$ Instituto de Nanociencia y Materiales de Aragón (INMA), CSIC-Universidad de Zaragoza, Zaragoza, 50009, Spain}

\date{\today}

\clearpage

\begin{abstract}
Defective chalcopyrites have recently emerged as promising thermoelectric materials because their ordered intrinsic vacancies can profoundly reshape both lattice dynamics and electronic structure. 
Here, we present a comprehensive theoretical investigation of the lattice thermal and carrier transport properties of II-III$_2$-VI$_4$ defective chalcopyrites by combining first-principles calculations with machine-learning interatomic potentials. 
We show that vacancy ordering enhances lattice distortion, leading to strong anharmonicity and metavalent bonding. 
The interplay of soft low-frequency phonons, strongly negative Gr\"uneisen parameters, and a substantially enlarged four-phonon-scattering phase space strongly regulates heat transport, yielding an ultralow lattice thermal conductivity.
Meanwhile, systematic anion substitution at the VI-site provides an effective route to tune the electronic structure: reduced anion electronegativity weakens metal-anion hybridization, shifts anion $p$ states upward, narrows the band gap, and thereby improves electrical transport. Benefiting from this synergy between vacancy-induced phonon suppression and anion-regulated electronic optimization, CdGa$_2$Te$_4$ exhibits an ultralow lattice thermal conductivity of 0.19 W$\cdot$m$^{-1}$K$^{-1}$ and a high room-temperature $ZT$ of 0.689. 
This work not only predicts defective chalcopyrites as a promising platform for high-performance thermoelectrics but also provides a practical design strategy by integrating vacancy ordering, higher-order phonon scattering, and anion-dependent band engineering.
\end{abstract}

\maketitle
\section*{Introduction}
Diamond-like compounds have emerged as a research hotspot in the field of thermoelectric materials due to their tetrahedrally coordinated crystal structures that can be traced back to binary zinc blende or wurtzite frameworks~\cite{JACS2025-LDZhao-Mercouri-diamondoid,Chen2025-diamondoid-review2025-acta-meta,kozai2025new,PRB2024-xiezhang-chalcopyrites,Angewan2022-wurtrite-Guilmeau,pang2022revealing,LeiHU-AdvEnergyMater2021}. 
Among them, representative systems with chalcopyrite (I-III-VI$_2$) have been widely investigated as well-known lattice-distorted thermoelectric materials~\cite{YiXIA-npj2026-cubic-tetra,PRB2026-JiangangHe-I-III-VI2,guo2025bidirectional,Small2023-ZHANG-I-III-VI2,cao2020origin,zhang2019design}. 
Their inherent lattice distortion induces strong lattice anharmonicity~\cite{li2022anharmonicity,zheng2022anharmonicity} and gives rise to ultralow lattice thermal conductivity ($\kappa_{L}$) that is a crucial foundation for high-performance thermoelectric materials~\cite{yue2023pressure,zheng2025ineffectiveness}.  
In addition, the electronic band structures are highly tunable in chalcopyrite compounds~\cite{wei2025pressure,miglio2017local}. 
For instance, substituting anionic components can effectively modulate the band gap, thereby optimizing electronic transport properties~\cite{miglio2017local,PRB2026-JiangangHe-I-III-VI2}.
To fully exploit the potential of these materials for applications, researchers have conducted extensive efforts in thermoelectric performance optimization in recent years through strategies such as element doping~\cite{JACS2026-Chalcopyrite-YiXIE,Mg-CuInTe2-ActaMater2024}, and external pressure~\cite{guo2025bidirectional,wei2025pressure}.

Although external stimuli can effectively improve thermoelectric performance, exploring compounds with intrinsic lattice distortion characteristics is crucial for fundamentally optimizing material properties~\cite{zhang2019evolution,ActaMater2023-defect-I-III-VI2-TiejunZHU,tan2026multifunctional}. 
Xie et al.~\cite{xie2022hidden} recently demonstrated that the large tetrahedral distortion parameter \(\eta\) (\(\eta = |2 - c/a|\)) in silver-based chalcopyrites drives their intrinsic low lattice thermal conductivity.
In-depth analysis reveals that this distortion induces strong lattice anharmonicity by breaking local tetrahedral symmetry, thereby significantly enhancing phonon scattering and ultimately suppressing the lattice thermal conductivity. 
Compared to traditional strategies that rely on external doping to introduce point defects for phonon scattering, intrinsic distortion achieves effective suppression of lattice thermal transport through the inherent characteristics of the crystal structure~\cite{caro2026phonon}, offering new possibilities for the development of a new generation of high-performance thermoelectric materials.

Defective chalcopyrites  (II-III$_2$-VI$_4$)  typically crystallize an ordered vacancy structure derived from the evolution of the chalcopyrite structure, achieved by doubling the original formula unit and removing one group-I atom, thereby creating vacancies at the original group-I atomic sites~\cite{xie2022hidden}. 
The chalcopyrite structure  ($I\bar{4}2d$, No.122) itself can be traced back to the ternary diamondoid compounds, as shown in Fig. 1(a).
It essentially comprises a double zinc blende cell ($F\bar{4}3m$, No.216)  stacked along the \textit{c}-axis, where monovalent and trivalent cations replace the divalent cation. 
In zinc blende, all tetrahedral coordination units are formed with perfect tetrahedral symmetry, and its structure can be characterized by a cubic lattice ($a = b = c$) and a unity tetragonal distortion ($\eta = 0$).
While the tetrahedral framework is preserved, it results in a tetragonal distortion characterized by $\eta\neq0$ and a deviation of bond angles from the ideal value ($109.47^\circ$), collectively manifesting the initial breaking of tetrahedral symmetry.
The further introduction of vacancies induces a second symmetry-lowering to $I\bar{4}$ (No. 82), thereby intensifying the lattice distortion.
As a result, the symmetry is reduced from  $F\bar{4}3m$ to  $I\overline{4}2d$, and then further decreases to $I\bar{4}$. 
Low-symmetry thermoelectric materials typically feature low $\kappa_L$ due to structure complexity.
Therefore, defective chalcopyrites also accommodate ultralow lattice thermal conductivity, with a magnitude comparable to the low $\kappa_L$ reported in other thermoelectric materials~\cite{1j9p-4wjv2025PRB,govindaraj2023ordered}.
This remarkable property renders it promising for thermoelectric applications, driving us to conduct systematic research on defective chalcopyrites.

In this work, we carry out a comprehensive investigation of the thermoelectric properties of defective chalcopyrites II-III$_2$-VI$_4$ by combining first-principles calculations with machine-learning interatomic potentials (MLIPs). We show that intrinsic vacancy ordering is not merely a crystallographic feature, but a key structural origin of the strong lattice anharmonicity in these materials. 
The vacancy-induced symmetry lowering and local lattice distortion foster metavalent bonding and soften low-frequency phonons, which collectively enhance higher-order phonon interactions.
This ultimately renders four-phonon scattering a key factor for suppressing lattice thermal conductivity.
In parallel, we reveal a clear anion-dependent electronic trend: as the electronegativity of the VI-site anion decreases, weakened orbital hybridization and the upward shift of anion $p$ states cooperatively narrow the band gap and improve electrical transport. As a result, CdGa$_2$Te$_4$ emerges as the best-performing member of this family, combining an ultralow lattice thermal conductivity of 0.19 W$\cdot$m$^{-1}$K$^{-1}$ with a room-temperature \textit{ZT} of 0.689. These findings clarify the coupled roles of vacancy ordering and anion engineering in regulating phonon and carrier transport, and provide a viable design strategy for high-performance chalcopyrite-based thermoelectric materials.

\section*{Results and Discussion}

\begin{figure*}[t]
\centering
\includegraphics[width=\textwidth]{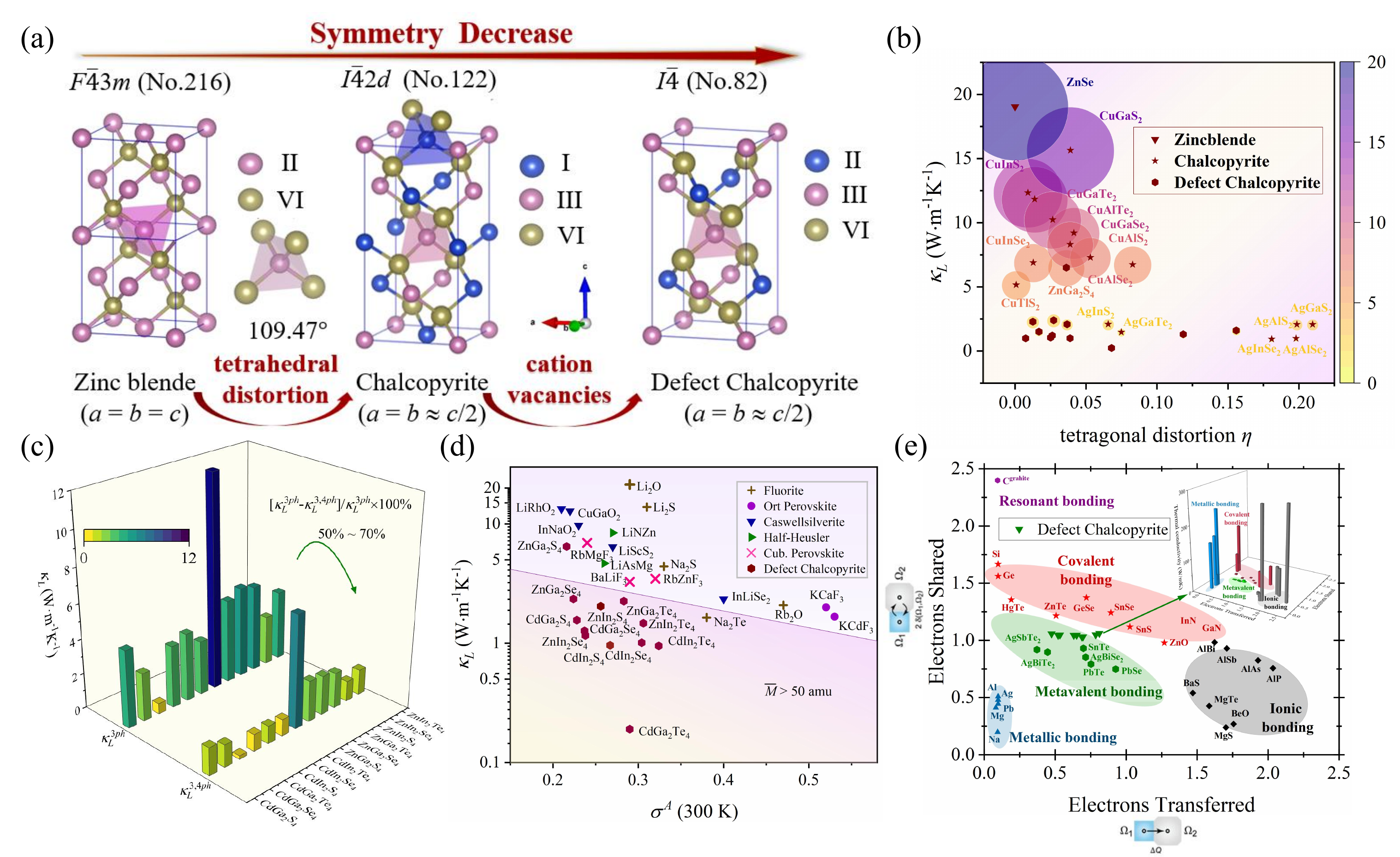}

\caption{
(a) The crystal structures of zinc blende, chalcopyrite, and defective chalcopyrites.
(b) The relationship between the $\eta$ and the $\kappa_L$ of different compounds.
The related data are available in Ref.~\cite{xie2022hidden}.
(c) The $\kappa_L$ evaluated by including both $\kappa_L^{3,4ph}$ and $\kappa_L^{3ph}$.
The color scale represents the $\kappa_L$.
(d) The $\kappa_L$ as a function of the lattice anharmonicity index $\sigma^A$ at 300 K.
The data for other compounds are given in Ref.~\cite{knoop2023anharmonicity}.
(e) The quantum mechanics diagram of bonding in solids.
The data for other compounds can be found in Ref.~\cite{raty2019quantum}.
}
\label{fig:1}
\end{figure*}

Before discussing the thermoelectric properties, we examined the mechanical and thermodynamic stability of the defective chalcopyrites.
Convex hull analysis based on the Open Quantum Materials Database (OQMD)~\cite{saal2013materials} in Fig. S8 confirms that the investigated compounds are thermodynamically stable. 
The results confirm that the investigated structures are stable, thereby providing a reliable basis for the subsequent thermoelectric analysis. The detailed stability criteria and calculated results are summarized in the Supporting Information.
As mentioned, the distortion parameter $\eta$ not only critically influences $\kappa_L$ but also serves as a fundamental indicator of lattice anharmonicity.
Generally, a larger $\eta$ indicates the greater deviation of the chalcopyrite lattice from the ideal diamond structure~\cite{miglio2017local}. 
Such a structural distortion directly reflects an increase in lattice anharmonicity, which in turn effectively suppresses lattice thermal conductivity. 
As shown in Fig. 1(b), a relatively high $\kappa_L$ is observed in the ideal cubic zinc blende structure ZnSe ($\eta = 0$).
As the lattice undergoes tetragonal distortion to chalcopyrite, the $\kappa_L$ significantly decreases. 
This trend clearly demonstrates that $\kappa_L$ is strongly suppressed as $\eta$ increases because of the enhanced lattice anharmonicity. 
Moreover, defective chalcopyrites possess lower-symmetry space groups and have more strongly distorted frameworks than chalcopyrite due to the presence of intrinsic cation vacancies. 
As expected, defective chalcopyrites promote intrinsically low lattice thermal conductivity, underscoring their potential as thermoelectric materials.

Chalcopyrites inherently feature pronounced phonon scattering driven by tetrahedral distortion, making higher-order processes indispensable for an accurate description~\cite{yue2023pressure,guo2025bidirectional}. 
Indeed, in CuInTe$_2$, the incorporation of 4\textit{ph} scattering results in a substantial reduction in $\kappa_L$ by 54.6\% at 800 K~\cite{yu2022temperature}. 
By contrast, defective chalcopyrites experience even stronger anharmonic scattering, arising from the cooperative effects of tetrahedral distortion and vacancy-induced local lattice perturbations. 
Fig. 1(c) shows the $\kappa_L$ (including 3\textit{ph} and 4\textit{ph} scattering) as a function of temperature for defective chalcopyrites. 
Our results reveal that the inclusion of 4\textit{ph} scattering leads to a 50-70\% reduction in lattice thermal conductivity in defective chalcopyrites at 300 K. 
Specifically, for CdGa$_2$Te$_4$, the inclusion of 4\textit{ph} scattering at 300 K reduces $\kappa_L$ from 0.61 W$\cdot$m$^{-1}$K$^{-1}$ (with only 3\textit{ph} scattering) to 0.19 W$\cdot$m$^{-1}$K$^{-1}$, corresponding to a reduction of 68\%. 
In addition, when only 3\textit{ph} scattering is considered, the calculated $\kappa_L$ of ZnIn$_2$Te$_4$ and CdIn$_2$Te$_4$ are 4.05 W$\cdot$m$^{-1}$K$^{-1}$ and 3.12 W$\cdot$m$^{-1}$K$^{-1}$ at 300 K, respectively. 
These values are substantially higher than the experimentally reported $\kappa_L$ of 2.1 W$\cdot$m$^{-1}$K$^{-1}$ for ZnIn$_2$Te$_4$ and 1.2 W$\cdot$m$^{-1}$K$^{-1}$ for CdIn$_2$Te$_4$~\cite{suriwong2011synthesis}, indicating that a description based solely on third-order anharmonicity overestimates $\kappa_L$.
Upon explicitly incorporating 4\textit{ph} scattering processes, the predicted $\kappa_L$ are significantly reduced to 1.47 W$\cdot$m$^{-1}$K$^{-1}$ and 0.95 W$\cdot$m$^{-1}$K$^{-1}$ for ZnIn$_2$Te$_4$ and CdIn$_2$Te$_4$, respectively, yielding much closer agreement with experiment. 
Notably, the remaining small deviation between our calculated and experimental values could also be influenced by factors such as sample quality and measurement accuracy, which are not accounted for in our ideal single-crystal model.
These results suggest that the cooperative effects of tetrahedral distortion and cation vacancies give rise to strong 4\textit{ph} scattering processes, whose contributions are essential for a reliable quantitative evaluation of $\kappa_L$ in defective chalcopyrites.

To quantify the strength of anharmonic effects in a material,  we calculate the anharmonicity index $\sigma^A(T)$, defined as:
\begin{equation}
    \sigma^{\mathrm{A}}(T) \equiv \frac{\sigma\left[F^{\mathrm{A}}\right]_{T}}{\sigma[F]_{T}}=\sqrt{\frac{\sum_{I, \alpha}\left\langle\left(F_{I, \alpha}^{\mathrm{A}}\right)^{2}\right\rangle_{T}}{\sum_{I, \alpha}\left\langle\left(F_{I, \alpha}\right)^{2}\right\rangle_{T}}} 
\end{equation}~\cite{knoop2020anharmonicity}.
A value of 
$\sigma^A(T) >$ 0.2 indicates that anharmonic interactions become comparable in magnitude to harmonic interactions~\cite{knoop2020anharmonicity}. 
As shown in Fig. 1(d), $\sigma^A(T = 300~\mathrm{K})$ values for defective chalcopyrites primarily lie in the range of 0.22-0.33. 
Among them, ZnGa$_2$S$_4$ exhibits the smallest $\sigma^A(300~\mathrm{K})$ and the largest $\kappa_L$ (6.46 W$\cdot$m$^{-1}$K$^{-1}$), whereas CdIn$_2$Te$_4$ shows the largest $\sigma^A(300~\mathrm{K})$ and a much lower $\kappa_L$ (0.95 W$\cdot$m$^{-1}$K$^{-1}$).
Extending the analysis to a broader range of materials with 
$\sigma^A(300~\mathrm{K})$ values from 0.2 to 0.6, including simple binary compounds, perovskites, and chalcopyrite-type systems, most materials are found to possess $\kappa_L$ below 10 W$\cdot$m$^{-1}$K$^{-1}$. 
Notably, defective chalcopyrites tend to show even lower $\kappa_L$ than other materials at comparable levels of anharmonicity.
This behavior can be largely attributed to their relatively high mean atomic mass ($\overline{M}$) and the presence of intrinsic vacancies.
Most of the defective chalcopyrites have an $\overline{M}$ greater than 50 amu (except ZnGa$_2$S$_4$ at 47.58 amu)~\cite{petley2002atomic}, while materials with higher $\kappa_L$ typically consist of lighter atoms.
These results suggest that the ultralow $\kappa_L$ in defective chalcopyrites arises from the synergistic effects of strong lattice anharmonicity, heavy atomic masses, and vacancy-induced phonon scattering.

\begin{figure*}[t]
\centering
\includegraphics[width=\textwidth]{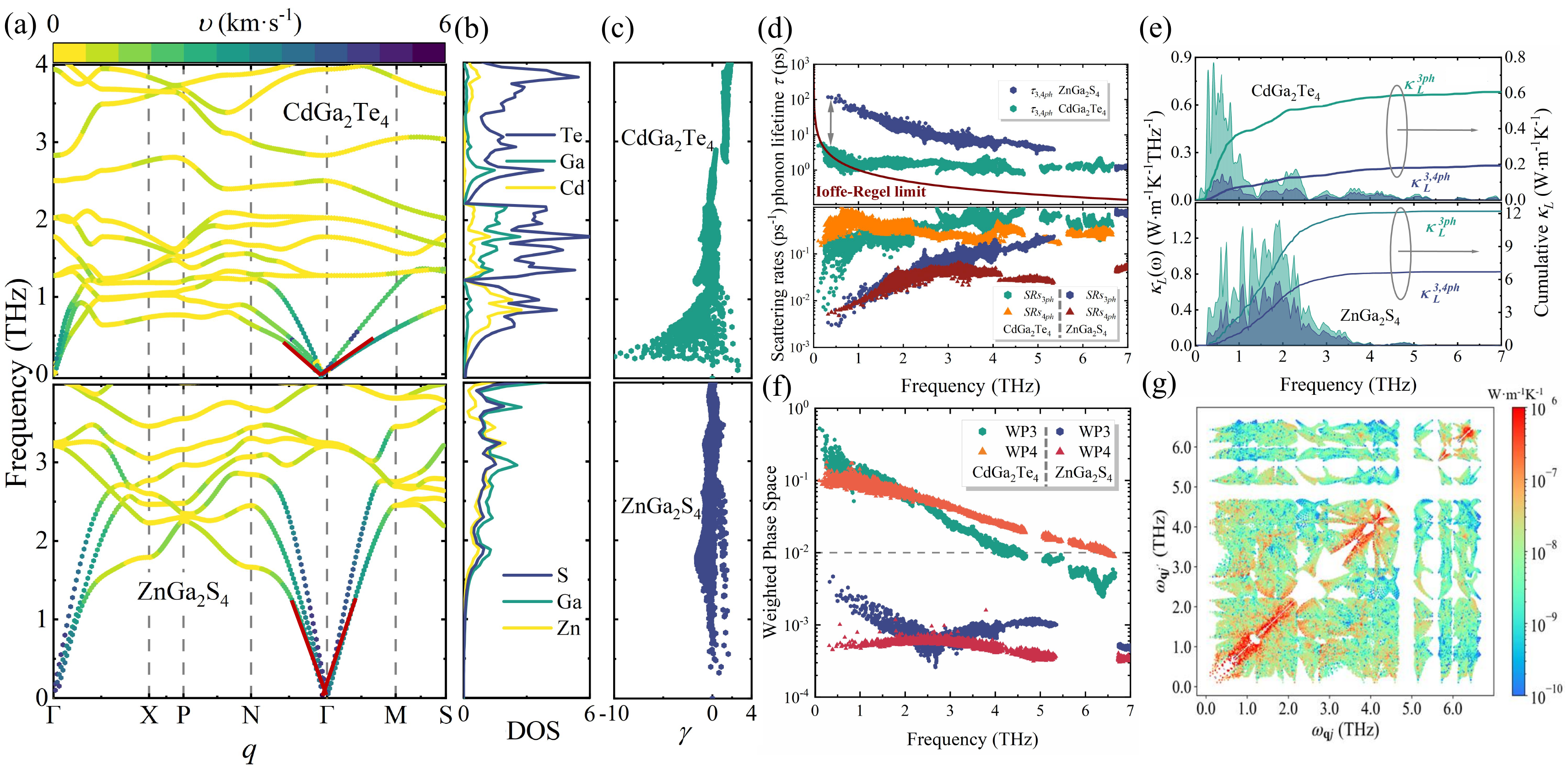}

\caption{
(a) Phonon dispersions in 0–4 THz. 
    (b) The phonon density of states.
    (c) The Grüneisen parameters $\gamma$ against the phonon frequency.
    (d) Phonon lifetimes $\tau_{3,4ph}$ including 3\textit{ph} and 4\textit{ph} scattering processes, as well as the scattering rates of 3\textit{ph} ($SRs_{3ph}$) and 4\textit{ph} ($SRs_{4ph}$), calculated at 300 K for CdGa$_2$Te$_4$ and ZnGa$_2$S$_4$. 
    (e) The spectral/cumulative lattice thermal conductivity at 300 K.
    (f) Three- and four-phonon weighted phase space at 300 K. 
    (g) The calculated $\kappa_C$($\omega_{qj}$,$\omega_{qj^{\prime}}$) for CdGa$_2$Te$_4$ at 300 K.
}
\label{fig:1}
\end{figure*}

On the other hand, intrinsic low lattice thermal conductivity is closely linked to the nature of its atomic bonding. 
Fig. 1(e) summarizes the charge transfer and the number of shared electron pairs based on atomic net charge $Q_{i}$ (reflecting electron transfer) and delocalization index $\delta(\Omega_{i},\Omega_{j})$ (quantifying covalent interactions between atoms)~\cite{yu2020chalcogenide}. 
These two descriptors clearly delineate bonding regimes associated with ionic, covalent, metallic, and metavalent bonding (MVB). 
Notably, defective chalcopyrites lie within the central green region, which corresponds to the MVB domain characterized by distinct transitional bonding behavior. 
Its bonding is characterized by approximately 0.8 e of shared electrons per bond, which is intermediate between conventional covalent bonding with highly localized electron sharing and metallic bonding with strongly delocalized electrons. 
Such unconventional bonding is considered the origin of its exceptionally low lattice thermal conductivity~\cite{yu2020chalcogenide,yue2024role}, as highlighted in the inset of Fig. 1(e). 
Detailed data for defective chalcopyrites are provided in Table S2 in the Supplementary Information.
Microscopically, this behavior arises from strong bond-induced lattice anharmonicity, manifested as a large Gr\"uneisen parameter~\cite{raty2019quantum}. 
Previous studies have shown that the intrinsically distorted tetrahedral configuration in silver-based chalcopyrite materials AgXY$_2$ promotes the formation of MVB~\cite{yuan2023soft}, leading to the softening of low-frequency optical phonons and the suppression of $\kappa_L$.
As a result, cation vacancies in defective chalcopyrites induce lattice distortion that promotes MVB and thereby introduces strong lattice anharmonicity, which ultimately leads to its extremely low lattice thermal conductivity.

For a comprehensive investigation of lattice thermal transport, CdGa$_2$Te$_4$ and ZnGa$_2$S$_4$ with the lowest and highest $\kappa_L$ were selected for comparison. 
As shown in Fig. 2(a), the calculated phonon dispersions of CdGa$_2$Te$_4$ and ZnGa$_2$S$_4$ do not show any imaginary modes, confirming their dynamical stability. 
The two compounds nevertheless differ markedly in the low-frequency region.
The ZnGa$_2$S$_4$ possesses a higher-lying lowest acoustic branch across the Brillouin zone, whereas softer acoustic phonons characterize CdGa$_2$Te$_4$.
To quantify the long-wavelength lattice vibrations, we linearly fitted the lowest acoustic branch along the $\Gamma-$M direction. 
The fitted slope for ZnGa$_2$S$_4$ is 2.04 THz$\cdot$\AA, which is substantially larger than the value of 0.61 THz$\cdot$\AA~obtained for CdGa$_2$Te$_4$, implying a higher acoustic phonon group velocity in ZnGa$_2$S$_4$.
This trend is mainly associated with the larger average atomic mass of CdGa$_2$Te$_4$, which lowers the vibrational frequencies and is expected to suppress the lattice thermal conductivity. 

Consistent with Fig. 2(a), the phonon density of states (DOS) in Fig. 2(b) shows that CdGa$_2$Te$_4$ is characterized by pronounced peaks dominated by Cd- and Te-derived vibrations, reflecting the dominant role of heavy atoms in the low-frequency modes. 
In contrast, ZnGa$_2$S$_4$ displays broader DOS features arising from the collective contributions of Zn, Ga, and S. Furthermore, we plotted the Gr\"uneisen parameter for CdGa$_2$Te$_4$ and ZnGa$_2$S$_4$ in Fig. 2(c). 
Within the 0-2 THz range, the Gr\"uneisen parameter of ZnGa$_2$S$_4$ stays close to zero and only reaches a mildly negative value of about $-$1.6, showing no pronounced peak and thus indicating relatively weak anharmonicity. 
On the other hand, a strongly negative Gr\"uneisen parameter appears in CdGa$_2$Te$_4$ over the same frequency window, with a pronounced minimum of around $-$10. 
This large negative value suggests that the presence of heavy atoms is closely linked to strong lattice anharmonicity, which in turn promotes enhanced higher-order phonon-phonon scattering.

Next, the influences of third- and fourth-order anharmonicity on phonon lifetime and scattering rates are investigated at 300 K. As seen from Fig. 2(d), CdGa$_2$Te$_4$ displays significantly shorter $\tau_{3,4ph}$ than ZnGa$_2$S$4$, particularly in the low-frequency region where $\tau_{3,4ph}$ is reduced by more than one order of magnitude. 
Such a difference in phonon lifetime is mainly attributed to the stronger lattice anharmonicity in CdGa$_2$Te$_4$, which enhances the scattering rates. The reduced group velocity in CdGa$_2$Te$_4$ does not affect the lifetime itself, but it indicates slower phonon propagation and thus less efficient heat transport, further suppressing the lattice thermal conductivity.

Within the frequency interval where the Gr\"uneisen parameter shows pronounced negative peaks, the low-frequency modes that are mainly associated with the heavy Cd and Te atoms experience stronger anharmonic interactions, which further shorten the phonon lifetime.
In addition, separating the three- and four-phonon contributions to the scattering rates reveals that, in the low-frequency range of CdGa$_2$Te$_4$, the 4\textit{ph} scattering rates are nearly an order of magnitude higher than those from three-phonon processes, indicating their significant role in suppressing the $\tau_{3,4ph}$. 
By contrast, ZnGa$_2$S$_4$ shows similar magnitudes of 3\textit{ph} and 4\textit{ph} scattering rates in the same low-frequency region, implying comparable contributions from the two scattering processes.
For both materials, 3\textit{ph} scattering rates become dominant above around 3.5~THz, particularly in ZnGa$_2$S$_4$. Taken together, the sharp reduction of phonon lifetime in CdGa$_2$Te$_4$ can be mainly attributed to the strongly enhanced higher-order scattering. In combination with its lower phonon group velocity, this leads to a stronger suppression of lattice thermal transport in CdGa$_2$Te$_4$ than in ZnGa$_2$S$_4$.

To further elucidate the significant role of 4\textit{ph} processes, the spectral and cumulative $\kappa_L$ shown in Fig. 2(e) are calculated with and without 4\textit{ph} scattering for comparison.
For CdGa$_2$Te$_4$, the reduction of $\kappa_L$ caused by 4\textit{ph} scattering is mainly located in 0-1.5 THz. 
The spectral heat-carrying contributions associated with 4\textit{ph }and 3\textit{ph} scattering show little overlap in this interval, which indicates that low-frequency heat transport is largely controlled by 4\textit{ph} scattering. 
By contrast, although 4\textit{ph} scattering in ZnGa$_2$S$_4$ also decreases $\kappa_L$ to some extent especially in 1-3 THz, its overall impact is weaker than in CdGa$_2$Te$_4$. 
Further insight into this contrast is obtained from the weighted phase space in Fig. 2(f). 
The 4\textit{ph} weighted phase space (WP4) of CdGa$_2$Te$_4$ is about two orders of magnitude larger than that of ZnGa$_2$S$_4$, which points to a larger number of available 4\textit{ph} scattering processes. 
Remarkably, WP4 is comparable to the 3\textit{ph} weighted phase space (WP3) in CdGa$_2$Te$_4$, so that 4\textit{ph} and 3\textit{ph} scattering share similar phase-space and higher-order processes can occur with high probability~\cite{han2022fourphonon}.
In ZnGa$_2$S$_4$, however, WP3 remains dominant in 0-2 THz. 
Only in 2-3 THz does WP4 become comparable to WP3, a trend that is consistent with the phonon spectrum of ZnGa$_2$S$_4$. 
Hence, in CdGa$_2$Te$_4$, the substantial WP4 and the pronounced intrinsic anharmonicity work together, where the former provides scattering processes, and the latter strengthens phonon-phonon coupling.
Together, they markedly enhance the scattering rate, thereby severely suppressing the lattice thermal conductivity. To examine whether the 4\textit{ph} scattering necessarily involves a wave-tunneling phonon transport mechanism~\cite{yue2024ultralow,yue2025diffuson}, we calculated the coherent lattice thermal conductivity of CdGa$_2$Te$_4$. As shown in Fig. 2(g) and Fig. S9, 
the results indicate that quasi-degenerate phonon modes $\left(\omega_{q j} \approx \omega_{q j'}\right)$ provide the main contribution to the coherent conductivity~\cite{zheng2024unravelling}. 
The maximum value of this contribution is only 10$^{-6}$ W$\cdot$m$^{-1}$K$^{-1}$, negligible compared with the total lattice thermal conductivity.

\begin{figure*}[t]
\centering
\includegraphics[width=\textwidth]{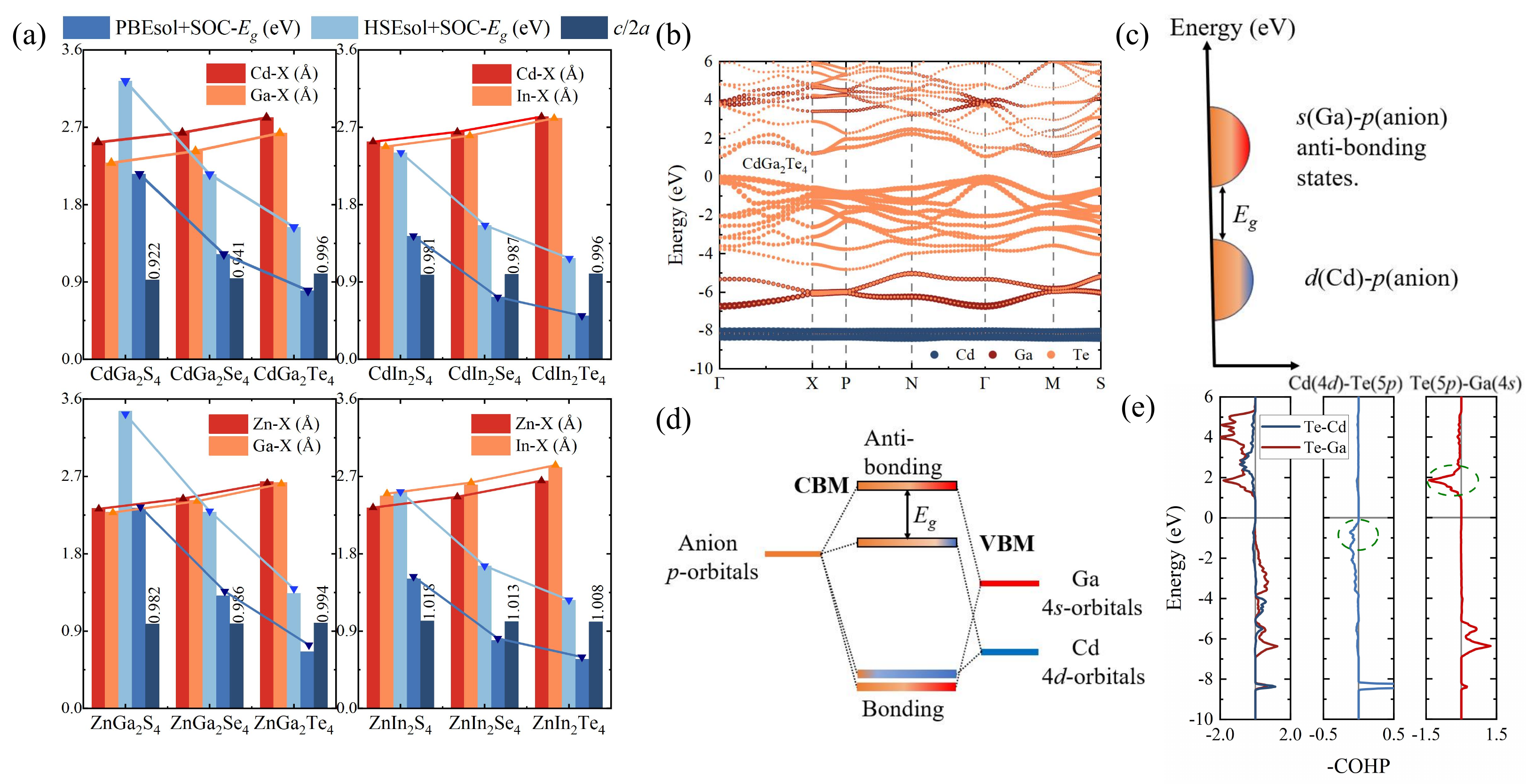}

\caption{
(a) Calculation results of bond lengths, structural parameter $c/2a$, and  electronic band gaps with PBEsol+SOC and HSEsol+SOC functionals for defective chalcopyrites. Detailed band gap data are provided in the Supplemental Material.
    (b) The atom-projected band structure of CdGa$_2$Te$_4$.
    (c) Schematic DOS with Cd-anion antibonding as the valence band edge and Ga-anion antibonding states as the conduction band edge.
    (d) Schematic molecular-orbital diagram.
    (e) Crystal Orbital Hamilton Population.
}
\label{fig:1}
\end{figure*}

As manifested in Fig. 3(a), there is a strong correlation between bond length, tetragonal distortion, and the band gap $E_g$ in defective chalcopyrites. 
For the CdGa$_2$X$_4$ (X = S, Se, Te), both Cd-X and Ga-X bond lengths increase monotonically  as the electronegativity of the VI-site anion decreases with the cation fixed, accompanied by an increase in 
$c/2a$ from 0.922 to 0.996, approaching the limit of $c/2a$ = 1. 
In parallel, the calculated $E_g$ decreases markedly with increasing bond length, and the same monotonic trend is also observed in the Zn-based counterparts. 
Since PBEsol tends to underestimate $E_g$, HSEsol results are included to provide a consistent comparison of the gap evolution. 
The above trend originates from the differences in the intrinsic properties of the anions.
With decreasing electronegativity of the VI-site anion and the cation fixed, the larger atomic radius weakens the cation-anion bonding, resulting in increased bond lengths and a gradual diminishment of tetragonal distortion.
These two factors collectively influence the electronic band structure, resulting in a monotonic decrease in the $E_g$ with increasing anion atomic number.

The band structure, atom-projected DOS, and bonding analysis for CdGa$_2$Te$_4$ were calculated to reveal the microscopic origin of the band gap reduction, as shown in Figs. 3(b)-(e). 
The band structure indicates a direct gap at $\Gamma$, with pronounced band degeneracy at the valence band maximum (VBM). 
As the electronegativity of the VI-site anion decreases with the cation fixed, the energy separation at the $\Gamma$ point between the valence bands near VBM gradually increases, while the band curvature near the Fermi level increases correspondingly (see Supplementary Figs. S10-S13 for details).
This increased curvature results in relatively steep band regions.
The projected DOS shows that the VBM is dominated by Te-5\textit{p} states, and the conduction band minimum (CBM) mainly originates from Ga-\textit{s} states with appreciable Te-\textit{p} admixture.
Cd-\textit{d} states are largely localized at deeper energies, yet their hybridization with anion \textit{p} states can contribute to the \textit{p}-\textit{d} repulsion that shapes the valence edge.
The bonding schematic in Fig. 3(d) provides an intuitive explanation for the aforementioned electronic structure features from a molecular orbital perspective. 
Ga(\textit{s})-Te(\textit{p}) coupling forms a bonding manifold in the deep valence region and an antibonding counterpart that constitutes the conduction-band edge. 
The Cd-Te interaction, through \textit{p}-\textit{d} orbital hybridization, forms bonding states within the valence band and generates antibonding states near the Te-5\textit{p} dominated VBM, which exhibit weak Cd-Te antibonding characteristics.
The Crystal Orbital Hamilton Population (COHP)~\cite{dronskowski1993crystal} in Fig. 3(e) further supports the finding by revealing antibonding characters associated with Ga-Te near the CBM and Cd-Te near the VBM. 
The trend in anion evolution further validates this mechanism.
 As the electronegativity of the VI-site anion decreases with the cation fixed, the metal-anion bonds become longer, and the hybridization weakens, which reduces the bonding-antibonding splitting; together with the upward shift of the anion-\textit{p} energy level for heavier chalcogens, this drives the VBM upward and tends to lower the antibonding CBM, resulting in an overall narrowing of $E_g$.

\begin{figure*}[t]
\centering
\includegraphics[width=\textwidth]{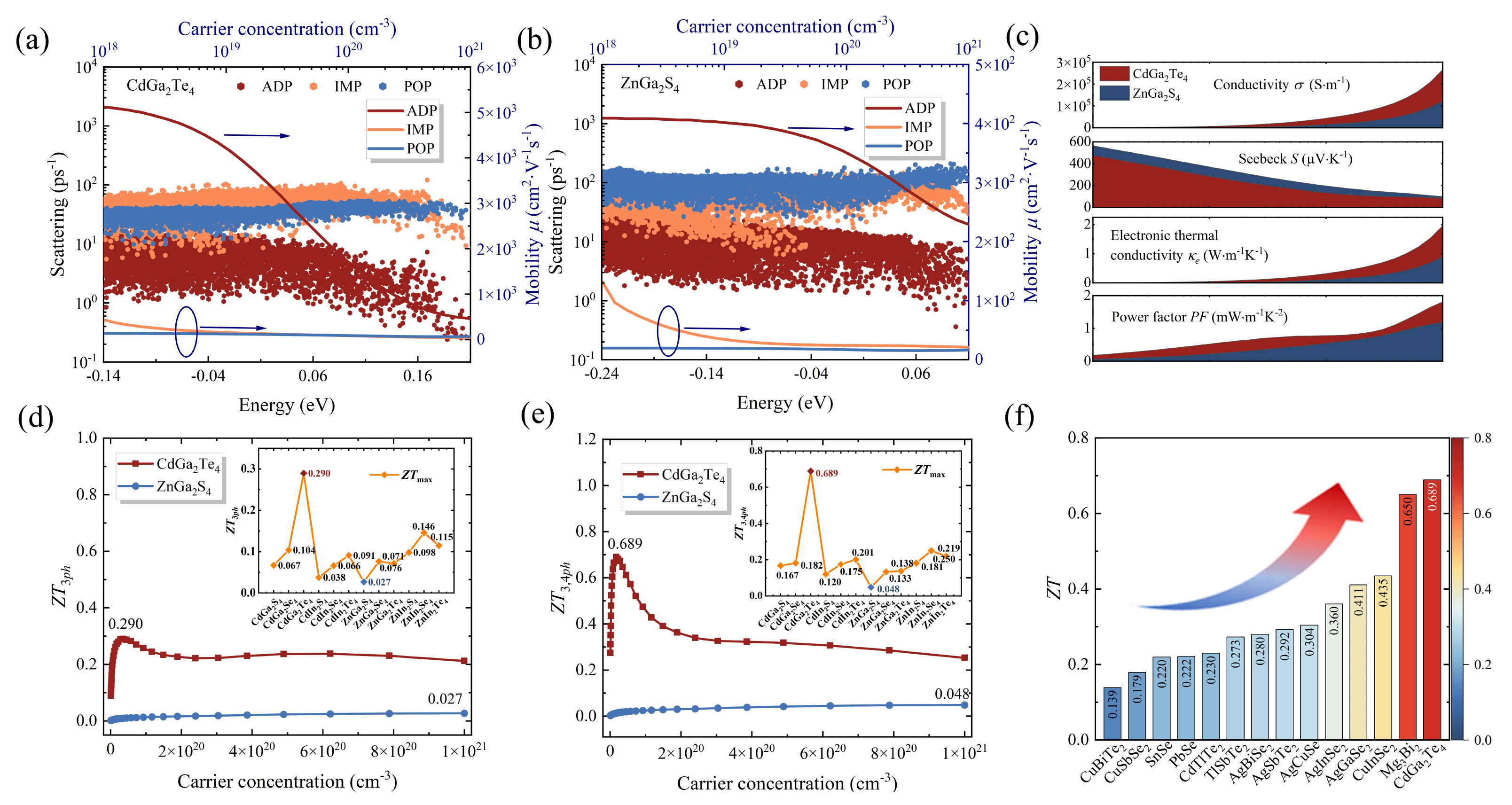}

\caption{
The scattering rates of ADP, IMP, and POP mechanisms for (a) CdGa$_2$Te$_4$ and (b) ZnGa$_2$S$_4$ at 300 K and a carrier concentration of $10^{21}$ cm$^{-3}$, and the corresponding carrier mobilities as functions of carrier concentration.
    (c) The calculated results of $\sigma$, $S$, $\kappa_e$, and $PF$ of CdGa$_2$Te$_4$ and ZnGa$_2$S$_4$ with varying carrier concentrations. 
The variation of (d) $ZT_{3ph}$ and (e) $ZT_{3,4ph}$ of CdGa$_2$Te$_4$ and ZnGa$_2$S$_4$ with carrier concentration considering.
(f) Comparison of $ZT$ between CdGa$_2$Te$_4$ and other typical thermoelectric compounds at 300 K, where the specific compounds corresponding to the labels are detailed in Refs.~\cite{chen2018high,liang2019flexible,li2024silver,wang2011reduction,shi2025crystal,liu2025enhanced,zhao2024plasticity}.
}
\label{fig:1}
\end{figure*}

The reduction in band gap $E_g$ typically lowers the energy threshold for electrons to be excited from the VBM to the CBM, leading to an increase in intrinsic carrier concentration and ultimately enhancing the electrical conductivity $\sigma$ as shown in Fig. S15.
The microscopic origin of the variation in $\sigma$ was further probed through a systematic analysis of the three dominant carrier-scattering mechanisms: ADP, IMP, and POP.
These mechanisms directly determine $\sigma$ by governing the carrier mobility $\mu$ ($\sigma=ne\mu$). 
The influence of scattering mechanisms on mobility follows a reciprocal relationship, i.e., carrier mobility is inversely proportional to the total scattering rate.
Therefore, a low scattering rate of a scattering mechanism indicates weaker hindrance to carrier motion, which is conducive to achieving higher mobility.
Figs. 4(a) and 4(b) present the variations in mobility for mechanisms as functions of carrier concentration and scattering rates in CdGa$_2$Te$_4$ and ZnGa$_2$S$_4$, respectively.
At 300 K and a carrier concentration of $10^{21}$ cm$^{-3}$, the scattering rate of ADP in both CdGa$_2$Te$_4$ and ZnGa$_2$S$_4$ is one order of magnitude lower than that of POP and IMP. 
Moreover, the scattering rates for mechanisms in CdGa$_2$Te$_4$ are slightly lower than those in ZnGa$_2$S$_4$. 
In both materials, the mobility dominated by ADP drops sharply when the carrier concentration exceeds $10^{19}$ cm$^{-3}$.
Remarkably, the mobility of CdGa$_2$Te$_4$  is one order of magnitude higher than that of ZnGa$_2$S$_4$, indicating its significantly higher $\sigma$.

Fig. 4(c) presents the carrier concentration dependence of $\sigma$, Seebeck coefficient $S$, electronic thermal conductivity $\kappa_e$, and power factor $PF$ for CdGa$_2$Te$_4$ and ZnGa$_2$S$_4$. As discussed earlier, CdGa$_2$Te$_4$ exhibits significantly higher $\sigma$ than ZnGa$_2$S$_4$. 
And this disparity becomes even more pronounced with increasing carrier concentration.
In contrast to the upward trend of $\sigma$, the $S$ of both materials gradually decreases as carrier concentration increases. 
ZnGa$_2$S$_4$ maintains a slightly higher $S$ than CdGa$_2$Te$_4$ overall, which is closely related to their differences in band structure. 
As shown in Fig. S11, ZnGa$_2$S$_4$ exhibits flatter band dispersion near the VBM, consistent with its larger hole effective masses along the $\Gamma$-N and $\Gamma$-M directions (0.766 $m_0$ and 0.578 $m_0$) compared to those of CdGa$_2$Te$_4$ (0.61 $m_0$ and 0.321 $m_0$). 
Since the $S$ is positively correlated with carrier effective mass, ZnGa$_2$S$_4$ consequently exhibits relatively higher $S$ values.
Notably, although ZnGa$_2$S$_4$ exhibits a slightly larger $S$ due to its heavier hole bands, CdGa$_2$Te$_4$ achieves a competitive $S$ through valence-band degeneracy, while its lighter bands ensure higher $\sigma$. This enables a synergistic optimization of $S$ and $\sigma$, resulting in an enhanced power factor. We also note the presence of conduction-band intervalley degeneracy in CdGa$_2$Te$_4$ (Fig.~S14), though it does not directly affect the present \textit{p}-type results.
Based on the above electronic transport parameters, we further calculated the evolution of the $PF$ with carrier concentration.
Despite the coupling relationship between $S$ and $\sigma$, CdGa$_2$Te$_4$ ultimately achieves a notable $PF$ owing to its considerably higher $\sigma$.
Additionally, the carrier concentration dependence of $\kappa_e$ follows essentially the same trend as $\sigma$, which is expected due to the Wiedemann-Franz law ($\kappa_e$ $= L\sigma$$T$, where $L$ is the Lorenz number)~\cite{chester1961law}. 
It is worth emphasizing that when evaluating the total thermal conductivity of these materials, the contribution from $\kappa_e$ cannot be neglected.
The  $\kappa_e$ is comparable to or even higher than that of $\kappa_L$ at elevated carrier concentrations.

Figs. 4(d) and 4(e) present the evolution of \textit{ZT} with carrier concentration for CdGa$_2$Te$_4$ and ZnGa$_2$S$_4$ at 300 K to evaluate the influence of 4\textit{ph} scattering by comparing the results under two scenarios: one considering only 3\textit{ph} scattering ($ZT_{3ph}$) and another incorporating both 3\textit{ph} and 4\textit{ph} scattering ($ZT_{3,4ph}$).
Benefiting from an ultralow $\kappa_L$ and a high \textit{PF}, CdGa$_2$Te$_4$ exhibits promising thermoelectric potential at 300 K. 
Its $ZT_{3ph}$ already reaches 0.270, being ten times larger than that of ZnGa$_2$S$_4$ (0.027). 
The inclusion of 4\textit{ph} scattering enhances \textit{ZT} for both materials, although the magnitude of the enhancement differs.
CdGa$_2$Te$_4$ shows a dramatic rise in value from 0.290 to 0.689, corresponding to a $\sim$2.4 times increase, whereas ZnGa$_2$S$_4$ exhibits only a modest increase from 0.027 to 0.048, marking a $\sim$1.8 times enhancement.
This reveals that 4\textit{ph} scattering plays a decisive role in defective chalcopyrites.
In addition, the \textit{ZT} increases monotonically as the electronegativity of the VI-site anion decreases with the cation fixed.
Such a trend aligns with the enhancement in $\sigma$ shown in Fig. S15.

Finally, we compared the $ZT$ of CdGa$_2$Te$_4$ with those of several representative thermoelectric materials at 300 K. 
As illustrated in Fig. 4(f), CdGa$_2$Te$_4$ achieves a notable $ZT$ of 0.689. 
This value drastically outperforms the other candidate materials such as the widely studied PbSe~\cite{liu2024efforts} and chalcopyrites (e.g., AgSbTe$_2$~\cite{li2024silver} and TlSbTe$_2$~\cite{wang2011reduction}), whose $ZT$ values generally fall below 0.45 at 300 K. 
The significant performance improvement demonstrates that defective chalcopyrites with intrinsic vacancies, particularly CdGa$_2$Te$_4$, exhibit immense potential as high-performance thermoelectric materials. 
This promising performance is mainly attributed to the strong lattice anharmonicity caused by its unique ordered intrinsic vacancy structure.
Such anharmonicity significantly enhances 4\textit{ph} scattering and reduces $\kappa_L$ to an ultra-low level. 
Meanwhile, the band structure ensures competitive electrical transport properties, enabling a favorable balance between thermal and electrical performance.
Our findings not only reveal the critical role of vacancies in regulating lattice thermal transport properties but also provide a new strategy for exploring next-generation high-performance thermoelectric materials.

\section*{Conclusion}

In summary, we have systematically elucidated the structure-property relationship governing thermoelectric performance in defective chalcopyrites. Ordered intrinsic vacancies are identified as the central structural motif responsible for their exceptional lattice thermal transport behavior. By further lowering crystal symmetry and amplifying local lattice distortion, vacancy ordering induces pronounced lattice anharmonicity, promotes metavalent bonding, and strengthens higher-order phonon interactions.
Together with soft low-frequency phonons, strongly negative Gr\"uneisen parameters, and reduced phonon group velocities, these effects make four-phonon scattering a major heat-resistance mechanism and drive the lattice thermal conductivity to ultralow values.
At the same time, the VI-site anion serves as an effective electronic approach: decreasing anion electronegativity elongates metal-anion bonds, weakens orbital hybridization, shifts anion $p$ states upward, and narrows the band gap, thereby enhancing carrier transport.
Owing to this synergistic optimization of phonon and electronic properties, CdGa$_2$Te$_4$ is identified as the optimal candidate in terms of overall thermoelectric performance, with a lattice thermal conductivity of 0.19 W$\cdot$m$^{-1}$K$^{-1}$ and a room-temperature $ZT$ of 0.689. Beyond identifying this high-performance thermoelectric compound, this work integrates vacancy ordering and anion engineering as a unified design principle for developing next-generation thermoelectric materials based on defective chalcopyrite frameworks.

\section*{Acknowledgement}

H.Z. and Y.L. acknowledge the National Key Research and Development Program of China (No. 2023YFA1406200) and the Institute of High-pressure Physics of Ningbo University for its computational resources. J.Y. acknowledges the China Scholarship Council (No. 202408330112). H.G. and Y.-W.F. acknowledge the financial support received from the IKUR Strategy under the collaboration agreement between Ikerbasque Foundation and Centro de F\'sica de Materiales (CFM-MPC) on behalf of the Department of Science, Universities and Innovation of the Basque Government (HPCAI21: AI-CrysPred). Y.-W.F. is partially supported by the Extraordinary Grant of CSIC (No. 2025ICT122) and acknowledge the support from the Spanish Ministry of Science and Innovation (PID2025-173011NA-I00) funded by MICIU/AEI/10.13039/501100011033. 
W.R. gratefully acknowledges the computing time provided by the NHR Center NHR4CES at RWTH Aachen University under project p0027528, with funding from the German Federal Ministry of Research, Technology and Space and the German federal states through the National High-Performance Computing (NHR) joint funding program.
S.L. acknowledges the computation resources provided by the National Academic Infrastructure for Supercomputing in Sweden (NAISS) -- at supercomputer centers NSC and PDC, partially funded by the Swedish Research Council through grant agreement No. 2022-06725 -- and by the Vienna Scientific Cluster (VSC) in Austria.

\section*{Conflict of Interest}

The authors declare no conflict of interest.

\section*{Author Contributions}

H.Z. and J.Y. contributed equally to this work. H.Z. and J.Y. conceived the project. J.Z. and H.G. performed code development. N.W., W.R., S.L., and C.S. tested and verified the data. H.Z., J.Y., and Y.-W.F. co-wrote the manuscript. Y.-W.F., Y.L., and T.C. supervised the research and acquired funding. All authors discussed the results and commented on the manuscript.

\section*{Data Availability Statement}

All data supporting this study's findings are available within the article. Additional data is available on request from the authors.

\bibliography{references}

@article{PRB2024-xiezhang-chalcopyrites,
  title = {Unveiling disparities and promises of {Cu and Ag} chalcopyrites for thermoelectrics},
  author = {Liang, Han-Pu and Geng, Songyuan and Jia, Tiantian and Li, Chuan-Nan and Xu, Xun and Zhang, Xie and Wei, Su-Huai},
  journal = {Phys. Rev. B},
  volume = {109},
  issue = {3},
  pages = {035205},
  numpages = {7},
  year = {2024},
  month = {Jan},
  publisher = {American Physical Society},
}

@article{YiXIA-npj2026-cubic-tetra,
  author = { Li, Zhi and Lee, Huiju and Wolverton, Chris and Xia, Yi },
  title = { High-throughput computational framework for high-order anharmonic thermal transport in cubic and tetragonal crystals },
  journal = {NPJ Comput Mater },
  year = { 2026 },
  volume = { 12 },
  pages = { 51 },
}

@article{Chen2025-diamondoid-review2025-acta-meta,
  author    = {Pengpeng Chen and Hongyao Xie and Li-Dong Zhao},
  title     = {Recent Progress on Diamondoid {Cu$_2$SnSe$_3$} Thermoelectric Materials: A Review},
  journal   = {Acta Metallurgica Sinica},
  year      = {2025},
  volume    = {38},
  number    = {5},
  pages     = {707--719},
}

@article{kozai2025new,
  title={A new quaternary sphalerite-derivative compound for thermoelectric applications: {Cu$_7$VSnS$_8$}},
author ="Kozai, Shuya and Suekuni, Koichiro and Takahashi, Seiya and Nishibori, Eiji and Kasai, Hidetaka and Siloi, Ilaria and Fornari, Marco and Saito, Hikaru and Sauerschnig, Philipp and Ohta, Michihiro and Lemoine, Pierric and Guilmeau, Emmanuel and Raveau, Bernard and Ohtaki, Michitaka",
journal  ="J. Mater. Chem. A",
year  ="2025",
volume  ="13",
issue  ="14",
pages  ="10028-10036",
publisher  ="The Royal Society of Chemistry",}

@article{JACS2025-LDZhao-Mercouri-diamondoid,
author = {Su, Jingyi and Liu, Yukun and Li, Yichen and Bai, Shulin and Gao, Dezheng and Chen, Pengpeng and Zhao, Zihao and Dravid, Vinayak P. and Xie, Hongyao and Zhao, Li-Dong and Kanatzidis, Mercouri G.},
title = {Intrinsic Off-Centering and Light Conduction Band Structure Lead to High Thermoelectric Performance in \textit{n}-Type Diamondoid {AgInSe$_2$}},
journal = {Journal of the American Chemical Society},
volume = {147},
number = {19},
pages = {16611-16619},
year = {2025},
}

@article{caro2026phonon,
  title={Phonon Scattering by Local Off-Centering in Diamond-Like {Cu$_{2-x}$Ag$_x$In$_2$Se$_4$} Chalcopyrites: High Carrier Mobility and Ultralow Thermal Conductivity},
  author={Caro-Campos, Irene and Posligua, Victor and Prado-Gonjal, Jes{\'u}s and Dura, Oscar J and Nemes, Norbert M and Gainza, Javier and Mart{\'\i}nez, Jos{\'e} L and Alonso, Jos{\'e} A and M{\'a}rquez, Antonio M and Plata, Jos{\'e} J and Serrano-Sanchez, Federico},
  journal={Small Structures},
  volume={7},
  number={2},
  pages={e202500800},
  year={2026},
  publisher={Wiley Online Library}
}

@article{LeiHU-AdvEnergyMater2021,
author = {Hu, Lei and Luo, Yubo and Fang, Yue-Wen and Qin, Feiyu and Cao, Xun and Xie, Hongyao and Liu, Jiawei and Dong, Jinfeng and Sanson, Andrea and Giarola, Marco and Tan, Xianyi and Zheng, Yun and Suwardi, Ady and Huang, Yizhong and Hippalgaonkar, Kedar and He, Jiaqing and Zhang, Wenqing and Xu, Jianwei and Yan, Qingyu and Kanatzidis, Mercouri G.},
title = {High Thermoelectric Performance through Crystal Symmetry Enhancement in Triply Doped Diamondoid Compound {Cu$_2$SnSe$_3$}},
journal = {Advanced Energy Materials},
volume = {11},
number = {42},
pages = {2100661},
year = {2021}
}

@article{Angewan2022-wurtrite-Guilmeau,
author = {Lemoine, Pierric and Guélou, Gabin and Raveau, Bernard and Guilmeau, Emmanuel},
title = {Crystal Structure Classification of Copper-Based Sulfides as a Tool for the Design of Inorganic Functional Materials},
journal = {Angewandte Chemie International Edition},
volume = {61},
number = {2},
pages = {e202108686},
year = {2022}
}

@article{pang2022revealing,
  title={Revealing an elusive metastable wurtzite {CuFeS$_2$} and the phase switching between wurtzite and chalcopyrite for thermoelectric thin films},
  author={Pang, Hong and Bourges, Cedric and Jha, Rajveer and Baba, Takahiro and Sato, Naoki and Kawamoto, Naoyuki and Baba, Tetsuya and Tsujii, Naohito and Mori, Takao},
  journal={Acta Materialia},
  volume={235},
  pages={118090},
  year={2022},
  publisher={Elsevier}
}

@article{zhang2019design,
author = {Zhang, Jian and Huang, Lulu and Zhu, Chen and Zhou, Chongjian and Jabar, Bushra and Li, Jimin and Zhu, Xiaoguang and Wang, Ling and Song, Chunjun and Xin, Hongxing and Li, Di and Qin, Xiaoying},
title = {Design of Domain Structure and Realization of Ultralow Thermal Conductivity for Record-High Thermoelectric Performance in Chalcopyrite},
journal = {Advanced Materials},
volume = {31},
number = {52},
pages = {1905210},
year = {2019}
}

@article{cao2020origin,
  title={Origin of the distinct thermoelectric transport properties of chalcopyrite {ABTe$_2$ (A= Cu, Ag; B= Ga, In)}},
  author={Cao, Yu and Su, Xianli and Meng, Fanchen and Bailey, Trevor P and Zhao, Jinggeng and Xie, Hongyao and He, Jian and Uher, Ctirad and Tang, Xinfeng},
  journal={Advanced Functional Materials},
  volume={30},
  number={51},
  pages={2005861},
  year={2020},
  publisher={Wiley Online Library}
}

@article{guo2025bidirectional,
  title={Bidirectional optimization of transport properties for thermoelectric performance via hydrostatic pressure in chalcopyrite {AgXTe$_2$ (X= In, Ga)}},
  author={Guo, Siqi and Yue, Jincheng and Zheng, Jiongzhi and Zhang, Hui and Wang, Ning and Li, Junda and Liu, Yanhui and Cui, Tian},
  journal={Physical Review B},
  volume={111},
  number={18},
  pages={184312},
  year={2025},
  publisher={APS}
}

@article{PRB2026-JiangangHe-I-III-VI2,
  title = {Thermoelectric properties of the copper-based chalcopyrite semiconductors {CuMX$_{2}$ (M = Al, Ga, and In; X = S, Se, and Te)} from first-principles calculations},
  author = {Xiong, Wu and Xia, Zhonghao and Han, Zhongjuan and Yao, Dong and He, Jiangang},
  journal = {Phys. Rev. B},
  volume = {113},
  issue = {7},
  pages = {075204},
  numpages = {16},
  year = {2026},
  month = {Feb},
  publisher = {American Physical Society},
}

@article{li2022anharmonicity,
  title={Anharmonicity-induced strong temperature-dependent thermal conductivity in {CuInX$_2$ (X= Se, Te)}},
  author={Li, Yongheng and Liu, Junyan and Hong, Jiawang},
  journal={Physical Review B},
  volume={106},
  number={9},
  pages={094317},
  year={2022},
  publisher={APS}
}

@article{yue2023pressure,
  title={Pressure-induced remarkable four-phonon interaction and enhanced thermoelectric conversion efficiency in {CuInTe$_2$}},
  author={Yue, Jincheng and Guo, Siqi and Li, Junda and Zhao, Jiahui and Shen, Chen and Zhang, Hongbin and Liu, Yanhui and Cui, Tian},
  journal={Materials Today Physics},
  volume={39},
  pages={101283},
  year={2023},
  publisher={Elsevier}
}

@article{miglio2017local,
  title={Local bonding influence on the band edge and band gap formation in quaternary chalcopyrites},
  author={Miglio, Anna and Heinrich, Christophe P and Tremel, Wolfgang and Hautier, Geoffroy and Zeier, Wolfgang G},
  journal={Advanced Science},
  volume={4},
  number={9},
  pages={1700080},
  year={2017},
  publisher={Wiley Online Library}
}

@article{wei2025pressure,
  title={Pressure-mediated band engineering and cation ordering for enhanced thermoelectric performance in {CuInTe$_2$} chalcopyrites},
  author={Wei, Jiaman and Yu, Wei and Li, Zhijian and Hu, Jiaxin and Wang, Tonghui and Yu, Yuan and Guo, Xin},
  journal={Chemical Engineering Journal},
  pages={166531},
  year={2025},
  publisher={Elsevier}
}

@article{Small2023-ZHANG-I-III-VI2,
author = {Huang, Lulu and Li, Yuanyue and Sha, Shengmao and Ge, Bangzhi and Wu, Yucheng and Yan, Jian and Kong, Yuan and Zhang, Jian},
title = {Engineering Multiple Microstructural Defects for Record-Breaking Thermoelectric Propertef Chalcopyrite {Cu$_{1-x}$Ag$_x$GaTe$_2$}},
journal = {Small},
volume = {19},
number = {15},
pages = {2206865},
year = {2023},
}

@article{ActaMater2023-defect-I-III-VI2-TiejunZHU,
title = {Achieving \textit{n}-type Conduction in {AMg$_2$Sb$_2$ (A=Yb, Eu, Ca, Sr, Ba)} {Zintl} Phases},
author={Zheng, Xin and Li, Airan and Liu, Shuo and Han, Zhongkang and Zhang, Min and Liu, Feng and Fu, Chenguang and Zhu, Tiejun},
journal = {Acta Materialia},
volume = {260},
pages = {119346},
year = {2023},
}

@article{tan2026multifunctional,
  title={Multifunctional Roles of Vacancy Defects in Advancing Thermoelectric Materials},
  author={Tan, Shuyue and Jiang, Lifeng and Kang, Huijun and Chen, Rongchun and Chen, Zongning and Guo, Enyu and Wang, Tongmin},
  journal={Small},
  pages={e13437},
  year={2026},
  publisher={Wiley Online Library}
}

@article{zhang2019evolution,
  title={Evolution of the intrinsic point defects in bismuth telluride-based thermoelectric materials},
  author={Zhang, Qi and Gu, Bingchuan and Wu, Yehao and Zhu, Tiejun and Fang, Teng and Yang, Yuxi and Liu, Jiandang and Ye, Bangjiao and Zhao, Xinbing},
  journal={ACS Applied Materials \& Interfaces},
  volume={11},
  number={44},
  pages={41424--41431},
  year={2019},
  publisher={ACS Publications}
}

@article{xie2022hidden,
  title={Hidden local symmetry breaking in silver diamondoid compounds is root cause of ultralow thermal conductivity},
  author={Xie, Hongyao and Bozin, Emil S and Li, Zhi and Abeykoon, Milinda and Banerjee, Soham and Male, James P and Snyder, G Jeffrey and Wolverton, Christopher and Billinge, Simon JL and Kanatzidis, Mercouri G},
  journal={Advanced Materials},
  volume={34},
  number={24},
  pages={2202255},
  year={2022},
  publisher={Wiley Online Library}
}

@article{knoop2023anharmonicity,
  title={Anharmonicity in thermal insulators: An analysis from first principles},
  author={Knoop, Florian and Purcell, Thomas AR and Scheffler, Matthias and Carbogno, Christian},
  journal={Physical Review Letters},
  volume={130},
  number={23},
  pages={236301},
  year={2023},
  publisher={APS}
}

@article{1j9p-4wjv2025PRB,
  title = {High-throughput screening of ordered vacancy compounds {AM$_{2}$X$_{4}$} for high-performance flexible thermoelectric materials},
  author = {Cheng, Linyuan and Li, Min and Zhang, Long and Wang, Hui},
  journal = {Phys. Rev. B},
  volume = {112},
  issue = {4},
  pages = {045202},
  numpages = {14},
  year = {2025},
  month = {Jul},
  publisher = {American Physical Society},
}

@article{govindaraj2023ordered,
  title={Ordered-vacancy defect chalcopyrite {ZnIn$_2$Te$_4$}: A potential thermoelectric material with low lattice thermal conductivity},
  author={Govindaraj, Prakash and Murugan, Kowsalya and Veluswamy, Pandiyarasan and Venugopal, Kathirvel},
  journal={Journal of Solid State Chemistry},
  volume={324},
  pages={124076},
  year={2023},
  publisher={Elsevier}
}

@article{Mg-CuInTe2-ActaMater2024,
title = {Realizing high thermoelectric performance and thermal stability in {CuInTe$_2$ through heavy dose Mg doping}},
author={Xiong, Qihong and Wu, Hong and Zhang, Kaiqi and Wang, Guiwen and Zheng, Sikang and Feng, Yajie and Wu, Shuai and Zhang, Bin and Han, Guang and Wang, Guoyu and others},
journal = {Acta Materialia},
volume = {278},
pages = {120268},
year = {2024},
}

@article{JACS2026-Chalcopyrite-YiXIE,
author = {Xu, Ting and Bai, Wei and Qin, Mi and Huang, LuLu and Liu, Yu and Wang, Shenghui and Li, Zhou and Yuan, Zhixiang and Xin, Hongxing and Li, Di and Huang, Zhulin and Chen, Ping and Liu, Xiaobing and Zhang, Yongsheng and Xiao, Chong and Zhang, Jian and Xie, Yi},
title = {Dual-Antisite Defects and Domain Structures Synergistically Boosting a Record-High {ZT $>$ 2.0 in Chalcopyrite Cu$_{0.7}$Ag$_{0.3}$Ga$_{1–x}$In$_x$Te$_2$(x = 0–0.5)}},
journal = {Journal of the American Chemical Society},
  volume={148},
  number={11},
  pages={12343--12353},
  year={2026},
  publisher={ACS Publications}
}

@article{saal2013materials,
  title={Materials design and discovery with high-throughput density functional theory: the open quantum materials database ({OQMD})},
  author={Saal, James E and Kirklin, Scott and Aykol, Muratahan and Meredig, Bryce and Wolverton, Christopher},
  journal={Jom},
  volume={65},
  number={11},
  pages={1501--1509},
  year={2013},
  publisher={Springer}
}

@article{han2022fourphonon,
  title={FourPhonon: An extension module to ShengBTE for computing four-phonon scattering rates and thermal conductivity},
  author={Han, Zherui and Yang, Xiaolong and Li, Wu and Feng, Tianli and Ruan, Xiulin},
  journal={Computer Physics Communications},
  volume={270},
  pages={108179},
  year={2022},
  publisher={Elsevier}
}

@article{yue2024ultralow,
  title={Ultralow glassy thermal conductivity and controllable, promising thermoelectric properties in crystalline {o-CsCu$_5$S$_3$}},
  author={Yue, Jincheng and Zheng, Jiongzhi and Li, Junda and Guo, Siqi and Ren, Wenling and Liu, Han and Liu, Yanhui and Cui, Tian},
  journal={ACS Applied Materials \& Interfaces},
  volume={16},
  number={16},
  pages={20597--20609},
  year={2024},
  publisher={ACS Publications}
}

@article{yue2025diffuson,
  title={Diffuson-Dominated Thermal Transport Crossover From Ordered to Liquid-Like {Cu$_3$BiS$_3$}: The Negligible Role of Ion Hopping},
  author={Yue, Jincheng and Zheng, Jiongzhi and Shen, Xingchen and Maji, Krishnendu and Yang, Chun-Chuen and Lin, Shuyao and Lemoine, Pierric and Guilmeau, Emmanuel and Liu, Yanhui and Cui, Tian},
  journal={Small},
  volume={21},
  number={42},
  pages={e06386},
  year={2025},
  publisher={Wiley Online Library}
}

@article{chester1961law,
  title={The law of Wiedemann and Franz},
  author={Chester, GV and Thellung, A},
  journal={Proceedings of the Physical Society},
  volume={77},
  number={5},
  pages={1005--1013},
  year={1961}
}

@article{raty2019quantum,
  title={A quantum-mechanical map for bonding and properties in solids},
  author={Raty, Jean-Yves and Schumacher, Mathias and Golub, Pavlo and Deringer, Volker L and Gatti, Carlo and Wuttig, Matthias},
  journal={Advanced Materials},
  volume={31},
  number={3},
  pages={1806280},
  year={2019},
  publisher={Wiley Online Library}
}

@article{suriwong2011synthesis,
  title={Synthesis and thermal conductivities of {ZnIn$_2$Te$_4$} and {CdIn$_2$Te$_4$} with defect-chalcopyrite structure},
  author={Suriwong, Tawat and Kurosaki, Ken and Thongtem},
  journal={Journal of alloys and compounds},
  volume={509},
  number={27},
  pages={7484--7487},
  year={2011},
  publisher={Elsevier}
}

@article{yu2022temperature,
  title={Temperature-dependent phonon anharmonicity and thermal transport in {CuInTe$_2$}},
  author={Yu, Hao and Chen, Liu-Cheng and Pang, Hong-Jie and Qiu, Peng-Fei and Peng, Qing and Chen, Xiao-Jia},
  journal={Physical Review B},
  volume={105},
  number={24},
  pages={245204},
  year={2022},
  publisher={APS}
}

@article{knoop2020anharmonicity,
  title={Anharmonicity measure for materials},
  author={Knoop, Florian and Purcell, Thomas AR and Scheffler, Matthias and Carbogno, Christian},
  journal={Physical Review Materials},
  volume={4},
  number={8},
  pages={083809},
  year={2020},
  publisher={APS}
}

@article{petley2002atomic,
  title={The atomic mass unit},
  author={Petley, BRIAN W},
  journal={IEEE Transactions on Instrumentation and Measurement},
  volume={38},
  number={2},
  pages={175--179},
  year={2002},
  publisher={IEEE}
}

@article{yu2020chalcogenide,
  title={Chalcogenide thermoelectrics empowered by an unconventional bonding mechanism},
  author={Yu, Yuan and Cagnoni, Matteo and Cojocaru-Mir{\'e}din, Oana and Wuttig, Matthias},
  journal={Advanced Functional Materials},
  volume={30},
  number={8},
  pages={1904862},
  year={2020},
  publisher={Wiley Online Library}
}

@article{yue2024role,
  title={Role of atypical temperature-responsive lattice thermal transport on the thermoelectric properties of antiperovskites {Mg$_3$XN (X= P, As, Sb, Bi)}},
  author={Yue, Jincheng and Liu, Yanhui and Ren, Wenling and Lin, Shuyao and Shen, Chen and Singh, Harish Kumar and Cui, Tian and Tadano, Terumasa and Zhang, Hongbin},
  journal={Materials Today Physics},
  volume={41},
  pages={101340},
  year={2024},
  publisher={Elsevier}
}

@article{yuan2023soft,
  title={Soft phonon modes lead to suppressed thermal conductivity in {Ag-based} chalcopyrites under high pressure},
  author={Yuan, Kunpeng and Zhang, Xiaoliang and Gao, Yufei and Tang, Dawei},
  journal={Physical Chemistry Chemical Physics},
  volume={25},
  number={36},
  pages={24883--24893},
  year={2023},
  publisher={Royal Society of Chemistry}
}

@article{dronskowski1993crystal,
  title={Crystal orbital Hamilton populations ({COHP}): energy-resolved visualization of chemical bonding in solids based on density-functional calculations},
  author={Dronskowski, Richard and Bloechl, Peter E},
  journal={The Journal of Physical Chemistry},
  volume={97},
  number={33},
  pages={8617--8624},
  year={1993},
  publisher={ACS Publications}
}

@article{liang2019flexible,
  title={Flexible thermoelectrics: from silver chalcogenides to full-inorganic devices},
author ={Liang, Jiasheng and Wang, Tuo and Qiu, Pengfei and Yang, Shiqi and Ming, Chen and Chen, Hongyi and Song, Qingfeng and Zhao, Kunpeng and Wei, Tian-Ran and Ren, Dudi and Sun, Yi-Yang and Shi, Xun and He, Jian and Chen, Lidong},
  journal={Energy \& Environmental Science},
  volume={12},
  number={10},
  pages={2983--2990},
  year={2019},
  publisher={Royal Society of Chemistry}
}

@article{chen2018high,
  title={High-performance {SnSe} thermoelectric materials: Progress and future challenge},
  author={Chen, Zhi-Gang and Shi, Xiaolei and Zhao, Li-Dong and Zou, Jin},
  journal={Progress in Materials Science},
  volume={97},
  pages={283--346},
  year={2018},
  publisher={Elsevier}
}

@article{li2024silver,
  title={Silver copper chalcogenide thermoelectrics: advance, controversy, and perspective},
  author={Li, Nan-Hai and Zhang, Qiang and Shi, Xiao-Lei and Jiang, Jun and Chen, Zhi-Gang},
  journal={Advanced Materials},
  volume={36},
  number={37},
  pages={2313146},
  year={2024},
  publisher={Wiley Online Library}
}

@article{liu2024efforts,
  title={Efforts Toward the Fabrication of Thermoelectric Cooling Module Based on \textit{n}-type and \textit{p}-type {PbTe} Ingots},
  author={Liu, Shibo and Qin, Yongxin and Wen, Yi and Shi, Haonan and Qin, Bingchao and Hong, Tao and Gao, Xiang and Cao, Qian and Chang, Cheng and Zhao, Li-Dong},
  journal={Advanced Functional Materials},
  volume={34},
  number={26},
  pages={2315707},
  year={2024},
  publisher={Wiley Online Library}
}

@article{wang2011reduction,
  title = {Reduction of thermal conductivity in PbTe:Tl by alloying with $\mathrm{TlSbT}{\mathrm{e}}_{2}$},
  author = {Wang, Heng and Charoenphakdee, Anek and Kurosaki, Ken and Yamanaka, Shinsuke and Snyder, G. Jeffrey},
  journal = {Phys. Rev. B},
  volume = {83},
  issue = {2},
  pages = {024303},
  numpages = {5},
  year = {2011},
  month = {Jan},
  publisher = {American Physical Society},
}

@article{zheng2022anharmonicity,
  title={Anharmonicity-induced phonon hardening and phonon transport enhancement in crystalline perovskite {BaZrO$_3$}},
  author={Zheng, Jiongzhi and Shi, Dongliang and Yang, Yuewang and Lin, Chongjia and Huang, He and Guo, Ruiqiang and Huang, Baoling},
  journal={Physical Review B},
  volume={105},
  number={22},
  pages={224303},
  year={2022},
  publisher={APS}
}

@article{zheng2024unravelling,
  title={Unravelling ultralow thermal conductivity in perovskite {Cs$_2$AgBiBr$_6$}: dominant wave-like phonon tunnelling and strong anharmonicity},
  author={Zheng, Jiongzhi and Lin, Changpeng and Lin, Chongjia and Hautier, Geoffroy and Guo, Ruiqiang and Huang, Baoling},
  journal={NPJ Computational Materials},
  volume={10},
  number={1},
  pages={30},
  year={2024},
  publisher={Nature Publishing Group UK London}
}

@article{zheng2025ineffectiveness,
  title={Ineffectiveness of formamidine in suppressing ultralow thermal conductivity in cubic hybrid perovskite {FAPbI$_3$}},
  author={Zheng, Jiongzhi and Chang, Zheng and Lin, Changpeng and Lin, Chongjia and Zhou, Yanguang and Huang, Baoling and Guo, Ruiqiang and Hautier, Geoffroy},
  journal={NPJ Computational Materials},
  volume={11},
  number={1},
  pages={315},
  year={2025},
  publisher={Nature Publishing Group UK London}
}

@article{liu2025enhanced,
  title={Enhanced thermoelectric performance in {p-type AgBiSe$_2$} through carrier concentration optimization and valence band modification},
  author={Liu, Hao-Ming and Wu, Xiu-Qun and Sun, Jia-Yan and Li, Shan and Zhang, Jun-Xiong and Ye, Xin-Li and Zhang, Qian},
  journal={Rare Metals},
  volume={44},
  number={2},
  pages={1213--1221},
  year={2025},
  publisher={Springer}
}

@article{zhao2024plasticity,
  title={Plasticity in single-crystalline {Mg$_3$Bi$_2$} thermoelectric material},
  author={Zhao, Peng and Xue, Wenhua and Zhang, Yue and Zhi, Shizhen and Ma, Xiaojing and Qiu, Jiamin and Zhang, Tianyu and Ye, Sheng and Mu, Huimin and Cheng, Jinxuan and others},
  journal={Nature},
  volume={631},
  number={8022},
  pages={777--782},
  year={2024},
  publisher={Nature Publishing Group UK London}
}

@article{shi2025crystal,
  title={Crystal symmetry modification enables high-ranged in-plane thermoelectric performance in {n-type SnSe} crystals},
  author={Shi, Haonan and Wen, Yi and Bai, Shulin and Chang, Cheng and Su, Lizhong and Gao, Tian and Liu, Shibo and Liu, Dongrui and Qin, Bingchao and Qin, Yongxin and others},
  journal={Nature Communications},
  volume={16},
  number={1},
  pages={1788},
  year={2025},
  publisher={Nature Publishing Group UK London}
}
\end{document}


\preprint{APS/123-QED}

\renewcommand{\thefigure}{S\arabic{figure}}
\setcounter{figure}{0}
\renewcommand{\thetable}{S\arabic{table}}
\setcounter{table}{0}

\title{Anharmonicity Driven by Vacancy Ordering Unlocks High-performance Thermoelectric Conversion in Defective Chalcopyrites II-III$_2$-VI$_4$}
\author{Hui Zhang$^{1,\dag}$, Jincheng Yue$^{2,3,1,\dag,*}$, Jiongzhi Zheng$^{4,5}$, Ning Wang$^{1}$, Wenling Ren$^{6}$, Shuyao Lin$^{7,8}$, Chen Shen$^{9,10}$, Hao Gao$^{3}$, Yanhui Liu$^{1,*}$, Yue-Wen Fang$^{2,*}$, Tian Cui$^{1,*}$} 
\affiliation{$^{1}$ Institute of High-Pressure Physics, School of Physical Science and Technology, Ningbo University, Ningbo 315211, China}

\affiliation{$^{2}$ Centro de Física de Materiales (CFM-MPC), CSIC-UPV/EHU, Manuel de Lardizabal Pasealekua 5, 20018 Donostia/San Sebastián, Spain}

\affiliation{$^3$ Fisika Aplikatua Saila, Gipuzkoako Ingeniaritza Eskola, University of the Basque Country (UPV/EHU), Europa Plaza 1, 20018 Donostia/San Sebastián, Spain}

\affiliation{$^4$ Energy Technologies Area, Lawrence Berkeley National Laboratory, Berkeley, CA 94720, USA}

\affiliation{$^5$ Thayer School of Engineering, Dartmouth College, Hanover, New Hampshire, 03755, USA}

\affiliation{$^6$ Institute of Materials Science, Technical University of Darmstadt, Alarich-Weiss-Strasse 2, Darmstadt, 64287, Germany}

\affiliation{$^7$ Technische Universit\"{a}t Wien, Institute of Materials Science and Technology, Vienna, A-1060, Austria}

\affiliation{$^8$ Link\"{o}ping University, Department of Physics, Chemistry, and Biology (IFM), Link\"{o}ping, SE-58183, Sweden}

\affiliation{$^9$ Suzhou Laboratory, Suzhou, China}

\affiliation{$^{10}$ Gusu Laboratory of Materials, Suzhou, China}

\date{\today}

       \maketitle

\begin{table}[htbp] 
\small
  \caption{\ The calculated elastic constants $C_{ij}$ (GPa), bulk modulus \textit{B} (GPa), Young’s modulus $\mathrm{\textit{E}}$ (GPa), shear modulus \textit{G} (GPa), and average wave velocity $\nu_{a}$ (m/s) for defective chalcopyrites II-III$_2$-VI$_4$.}
  \label{tbl:example2}
  \begin{tabular*}{\textwidth}{@{\extracolsep{\fill}}cccccccccccc}
\hline
& \multicolumn{1}{c}{$C_1$$_1$} & \multicolumn{1}{c}{$C_1$$_2$} & \multicolumn{1}{c}{$C_1$$_3$} & \multicolumn{1}{c}{$C_3$$_3$}& \multicolumn{1}{c}{$C_4$$_4$} &\textit{B} & \textit{E} & \textit{G} &  $\nu_a$ \\
\hline
CdGa$_2$S$_4$ & 58.818 & 27.888 & 34.063 & 60.286& 34.068 &  41.014 & 57.868 & 22.876 &   2638.703 \\
CdGa$_2$Se$_4$ & 52.853 & 23.476 & 30.992 &  53.488&  30.021& 36.555 & 50.913 & 20.078 &  2161.905 \\
CdGa$_2$Te$_4$ & 44.170 & 19.007 & 25.400 & 47.305 &  25.225& 30.405 & 43.789 & 17.377 &  1932.009 \\
CdIn$_2$S$_4$ & 40.032 & 24.271 & 24.839 &44.618  & 24.299 & 30.232 & 40.671 & 15.940 &   2114.863 \\
CdIn$_2$Se$_4$ & 39.038 & 22.800 & 24.039 &44.305 & 22.644 & 29.262 & 38.993 & 15.257 &  1850.390 \\
CdIn$_2$Te$_4$ & 36.656 & 19.229 & 21.173 &39.455 & 20.224 &26.163 & 36.136 & 14.229 &   1734.300 \\
ZnGa$_2$S$_4$ & 58.818 & 27.888 & 34.063 &60.286 & 34.068 &41.014 & 57.868 & 22.876 &  2638.703 \\
ZnGa$_2$Se$_4$ & 62.885 & 29.834 & 33.074 &66.748&  36.165& 42.660 & 64.902 & 26.035 &   2473.327 \\
ZnGa$_2$Te$_4$ & 52.314 & 23.948 & 26.180 & 54.773&  28.820&   34.638 & 53.259 & 21.411 & 2145.134 \\
ZnIn$_2$S$_4$ & 50.603 & 32.338 & 27.587 &53.814  &  27.638&36.668 & 50.475 & 50.475 & 2395.067  \\
ZnIn$_2$Se$_4$ & 48.416 & 28.876 & 25.893 & 49.482 &25.043 & 34.176 & 47.218 & 18.594 &  2052.072 \\
ZnIn$_2$Te$_4$ & 43.276 & 23.968 & 21.544 &42.427& 21.576 & 29.214 & 41.801 & 16.568 &   1873.700 \\
       \hline
  \end{tabular*}
\end{table}

\begin{figure}
    \centering
    \includegraphics[width=1\linewidth]{FigureS2.pdf}
    \caption{Comparison in energies between the moment tensor potentials (MTPs) model and density functional theory (DFT) for all compounds.}
    \label{fig:S1}
\end{figure}

\begin{figure}
    \centering
    \includegraphics[width=1\linewidth]{FigureS3.pdf}
    \caption{Comparison in forces between the MTP model and DFT for all compounds.}
    \label{fig:S2}
\end{figure}

\begin{figure}
    \centering
    \includegraphics[width=1\linewidth]{FigureS4.pdf}
    \caption{Comparison in energies between the Deepmd (DP) model and DFT for all compounds.
    }
    \label{fig:S3}
\end{figure}

\begin{figure}
    \centering
    \includegraphics[width=1\linewidth]{FigureS5.pdf}
    \caption{Comparison in forces between the DP model and DFT for all compounds.}
    \label{fig:S4}
\end{figure}

\begin{figure}
    \centering
    \includegraphics[width=1.0\linewidth]{FigureS6.pdf}
    \caption{Molecular dynamics simulations in the canonical ensemble calculated using the DP model with a total simulation time of 100 ps. The green curve represents the potential energy, and the blue curve represents the thermostat dynamics.}
    \label{fig:S5}
\end{figure}

\begin{figure}
    \centering
    \includegraphics[width=1\linewidth]{FigureS7.pdf}
    \caption{Convergence tests for thermal conductivity. (a) Calculated thermal conductivity as a function of the neighbor-shell cutoff using third-order interatomic force constants (IFCs); (b) Calculated thermal conductivity as a function of the \textit{q}-mesh density including fourth-order IFCs.}
    \label{fig:S6}
\end{figure}

\begin{figure}
    \centering
    \includegraphics[width=1.0\linewidth]{FigureS8.pdf}
    \caption{Error distribution of atomic forces obtained from MTP-based potentials compared to DFT calculations. The horizontal axis represents the force error (eV$\cdot$\AA$^{-1}$), and the vertical axis represents the probability density. The RMSE of the forces for all compounds is less than 0.002 eV$\cdot$\AA$^{-1}$.}
    \label{fig:S7}
\end{figure}

\begin{figure}
    \centering
    \includegraphics[width=1.0\linewidth]{FigureS1.pdf}
    \caption{The convex hull diagrams for assessing the thermodynamic stability of all compounds were obtained from the Open Quantum Materials Database (OQMD). Any compound on the convex hull (0 meV$\cdot$atom$^{-1}$ relative to the convex hull) is thermodynamically stable.}
    \label{fig:S8}
\end{figure}

\begin{figure}
    \centering
    \includegraphics[width=0.6\linewidth]{FigureS9.pdf}
    \caption{Coherent thermal conductivity $\kappa_C$ of all compounds.}
    \label{fig:S9}
\end{figure}

\begin{figure}
    \centering
    \includegraphics[width=0.95\linewidth]{FigureS10.pdf}
    \caption{Schematic diagram of the band structure. The solid black line connects the conduction band minimum (CBM) and the valence band maximum (VBM), with the band gap $E_g$ labeled alongside. The Fermi level is set to zero energy.}
    \label{fig:S10}
\end{figure}

\begin{figure}
    \centering
    \includegraphics[width=1\linewidth]{FigureS11.pdf}
    \caption{Calculated Crystal Orbital Hamilton Population (COHP) for all compounds, along with the orbital-projected DOS distributions for different atoms.}
    \label{fig:S11}
\end{figure}

\begin{figure}
    \centering
    \includegraphics[width=1\linewidth]{FigureS12.pdf}
    \caption{Calculated total and orbital-projected density of states (DOS) distributions for all compounds.}
    \label{fig:S12}
\end{figure}

\begin{figure}
    \centering
    \includegraphics[width=0.9\linewidth]{FigureS13.pdf}
    \caption{Calculated carrier-concentration-dependence of (a) electrical conductivity $\sigma$, (b) Seebeck coefficient $S$, (c) electronic thermal conductivity $\kappa_e$, and (d) power factor $PF$.}
    \label{fig:S13}
\end{figure}

\begin{figure}
    \centering
    \includegraphics[width=1\linewidth]{FigureS14.pdf}
    \caption{Calculated heatmap comparison of the thermoelectric figure of merit \textit{$ZT$} considering only three-phonon scattering (\textit{$ZT_{3ph}$}) and including both three- and four-phonon scattering (\textit{$ZT_{3,4ph}$}) as a function of carrier concentration.}
    \label{fig:S14}
\end{figure}